**Assessment of nacre-like ceramics in replacement to Ni superalloys in aircraft's engines**

Jie Sheng Chan[1], Hortense Le Ferrand[1,2]

[1]School of Mechanical and Aerospace Engineering, 50 Nanyang Avenue, Nanyang Technological University, 639798 Singapore

[2]School of Materials Science and Engineering, 50 Nanyang Avenue, Nanyang Technological University, 639798 Singapore

**Corresponding e-mail:** hortense@ntu.edu.sg

**Abstract**

Aviation's fossil fuel emissions contribute to global warming. The production and disposal of the materials used in aircrafts too. The current metallic alloys present in the hot section of engines pose constraints in terms of temperature, pressure and weight that restrain the performance of the aircrafts. Also, these alloys are produced using rare, depleting resources, and polluting processes. In this paper, we hypothesize the use of bioinspired nacre-like alumina (NLA), a ceramic material that exhibits unusual toughness, and evaluate its potential as a replacement for superalloys in aircraft's engines. Comparing the performance of Ni superalloys and NLA in terms of properties, engine performance, and life cycle sustainability, we find NLA a promising alternative although progress has to be made with regards to its reliability, shaping, repair, and governance of the production process.

# 1| Introduction

Global warming and environmental pollution are today's and tomorrow's major challenges. Although aviation contributed to only 0.8% of the total anthropogenic warming associated to fossil fuel emissions in 2000, its emissions and global warming potential is expected to at least double by 2050, even if the Paris Agreement is followed [1]. Indeed, the traffic rate from 2015 to 2034 is predicted to increase by 4 to 5% per year [1]. Aviation is detrimental to the environment through the emission of greenhouse gases such as $CO_2$, NOx, SOx and of soot, mainly. Furthermore, aviation produces other types of pollution such as noise, destruction of habitat, usage of water, fuel, electricity, etc. In the context of the sustainable development goals established by the United Nations for 2030, it is important to evaluate aviation's environmental impact beyond greenhouse gases emitted during flight and to consider the entire life cycle, comprising environmental, economic and social impacts, from the fabrication of the raw materials to the disposal of the engines [2,3].

Among the proposed approaches to mitigate aircrafts' impacts such as the optimization of the aerodynamics, flight management, and the use of biofuels, reducing the aircraft's weight has already had successful outcomes [4]. This weight reduction has been allowed mostly by replacing metallic parts with composite materials such as carbon fiber reinforced polymers (CFRPs) or hybrid glass aluminum reinforced epoxy laminates (GLARE) [5], and has been reported to decrease the overall environmental impact of aircrafts during their flight [2,6,7]. However, with the continuous increase in air traffic, there is still progress to be made to further reduce aviation-induced pollution. To attain a carbon neutral status, the International Civil Aviation Organization (ICAO), has laid out a comprehensive framework in which improving propulsion technology is one strategy [8]. Indeed, improving the efficiency of the engines could have a significant contribution in reducing the amount of fuel required and exhaust gas emissions. Increasing engine's efficiency could be achieved by using more lightweight, temperature resistant materials that increase the temperature and pressure ratios in the turbine. Currently, nickel-based single crystal superalloys CSMX-10 are the most commonly used in the last generation of turbine blades [9]. However, thermal oxidation and corrosion limit their operating temperatures and protective coatings and cooling channels inside the blades are required, making their design complex and costly [10]. These protective coatings, although efficient, are made of ceramics and are prone to spalling, increasing the risk for crack initiation and growth, whereas the air flow within the cooling channels limits the overall efficiency of the turbine [11,12]. To remediate to these drawbacks, SiC/SiC ceramic matrix composites (CMCs) have been developed using melt infiltration processes [11]. Yet, SiC turbine blades are still sensitive to high temperature water vapor and their processing requires a manual layup process. [9] Current research efforts now tackle the development of new alloy compositions, such as high entropy alloys or Nd-silicide alloys, among others [13,14].

In the quest for lightweight temperature-resistant materials, bioinspired refractory ceramics present interesting advantages. Indeed, research of the past decades has uncovered some of the structural designs of strong and tough natural ceramic composites such as the dactyl club of the Mantis Shrimp [15], the multi layered arrangements of the Conch shell [16], and the nacreous layers of seashells [17,18]. In particular, the study of nacre has led to the development of biomimetic or bioinspired materials that reproduce the key toughening mechanisms without compromising on strength. Taking inspiration from nacre, the brick and mortar aligned arrangement of 95% aragonite ($CaCO_3$) platelets found in seashells (**Figure 1A**) has been reproduced in technical ceramic systems using $Al_2O_3$ microplatelets (**Figure 1B**). Thanks to the brick and mortar organization, crack deflection, crack bridging, and friction at grain interfaces allow energy dissipation and increase toughness without diminishing strength (**Figure 1C**). Contrary to SiC, $Al_2O_3$ can withstand high temperatures in humid and corrosive atmospheres. These properties make nacre-like alumina (NLA) a potential replacement for lightweight turbine blades application in aerospace engines. Although NLAs are still currently researched to further improve their properties, investigate their behavior, and establish the appropriate fabrication methods, the results obtained thus far calls for consideration and assessment towards their future use in industrial settings. Such consideration needs to be aligned with the sustainable development goals to anticipate, estimate benefits and prevent potential negative impacts.

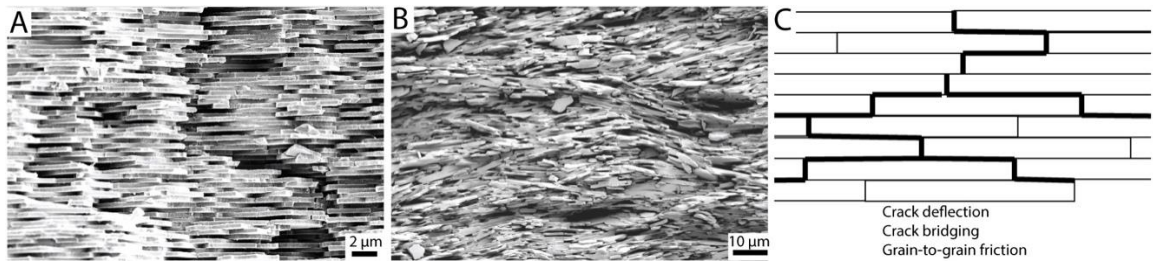

**Figure 1:** **(A)** Electron micrograph of the fractured natural nacre from an Abalone seashell (image courtesy of T. Niebel). **(B)** Electron micrograph of a nacre-like alumina (image from the authors). **(C)** Schematics of key toughening mechanisms occurring in nacre-like brick and mortar microstructures.

In this paper, we perform a thought experiment where NLAs are used as turbine blade material in engines for aircraft propulsion and assess their advantages and drawbacks using data reported in the literature. We therefore make the hypothesis that NLAs can be fabricated in shapes and scales of turbine blades, despite an example of such is unavailable as of today. To draw meaningful comparisons, the performance of NLAs in terms of materials properties, environmental and socio-economic impacts are compared with the standard, in-service nickel-based superalloys (**Figure 2**). First, a materials properties comparison is carried out to discuss the engine performance. Next, a socio-economic-environmental evaluation is drawn at each phase of the life cycle of the hypothetic turbine blade, *i.e.* mining, fabrication, and end of life. The outcomes of the analysis are then discussed to identify remaining challenges and areas of improvement, to better guide research and development in this area and to excite stakeholders with this new material.

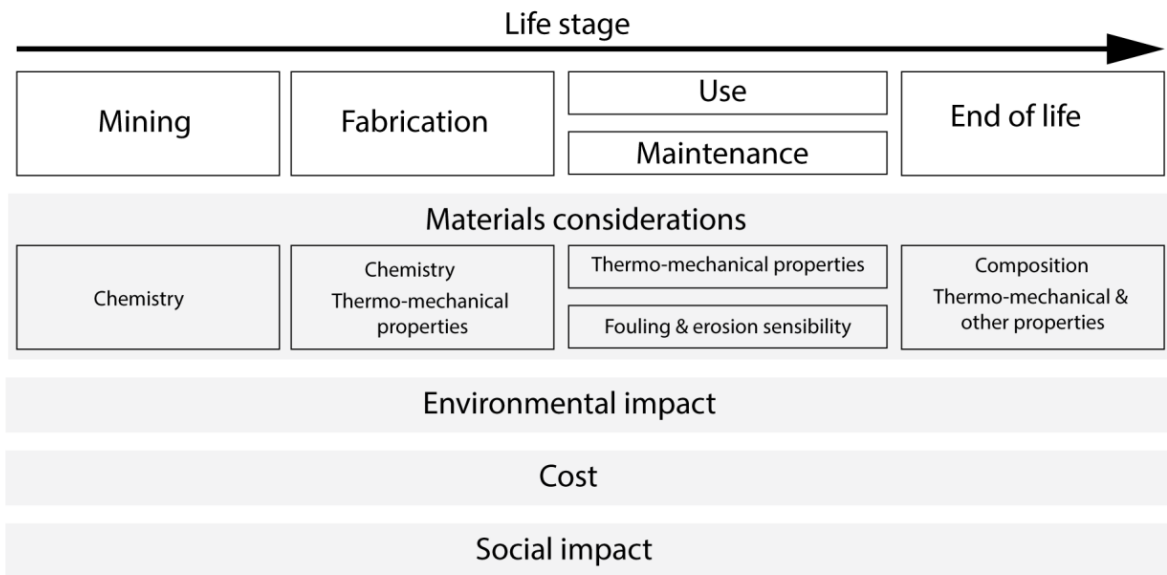

**Figure 2:** Schematic overview of the areas considered in this paper for the assessment of turbine blades made of NLA.

**2| Methodology**

**2.1| System considered**

The materials considered are the NLA produced following Bouville et al [19] and prepared using the casting method as per Hassan et al [20]. The preparation method in these two papers led to similar properties but the latest one presents a simpler and faster route. Latest generation Ni superalloy CMSX-10 is used for comparison when the data are available. Other data were extracted from publications using CMSX-4 and other similar Ni superalloys. The chemical composition and processing methods are presented in **Table 1** and **Figure 3**, respectively.

**Table 1**: Composition of CMSX-10 superalloy[21] and of NLA.[19]

|          | CMSX-10 |    |       |      |   |       |    |     |    |      |    | NLA |      |     |
|----------|---------|----|-------|------|---|-------|----|-----|----|------|----|-----|------|-----|
| Material | Cr | Co | Ni | Mo | W | Nd/Cb | Ta | Ti | Al | Hf | Re | $Al_2O_3$ | $SiO_2$ | CaO |
| Wt%      | 2  | 3  | 69.57 | 0.40 | 5 | 0.2 | 8 | 0.2 | 5.7 | 0.03 | 6 | 98.5 | 1.3 | 0.2 |

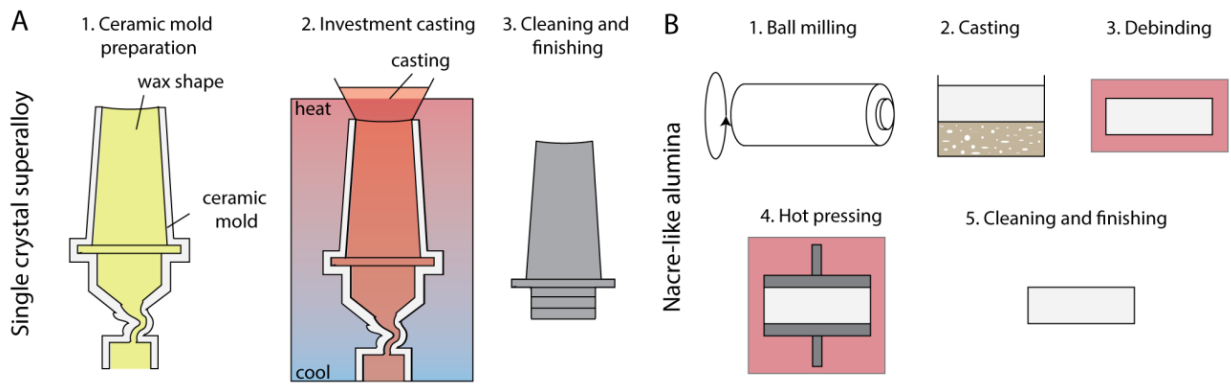

**Figure 3:** Main steps for superalloy and NLA fabrication.

**2.2| Data gathering**

Materials related data were gathered from the literature using CMSX-10 Ni superalloys, CMSX-4 Ni superalloys, or, when not available, other standard Ni superalloys. Similarly, for NLA, the data were gathered from Bouville et al and Hassan et al when possible. For high temperature properties, the NLAs developed by Pelissary et al [22] were used. Indeed, Bouville et al NLAs contain a silica mortar allowing testing up 600 °C [19], whereas Pelissary et al. NLAs were free of mortar and could be tested at 1200 °C [22]. Other alumina-related properties were taken from traditional alumina. The data presented are averages. For environmental, economic and social impacts, the main components of Ni superalloys and NLA were considered only, namely nickel and alumina. Other data were accumulated from the appropriate literature. Transportation and storage of goods were not considered in the assessment. Details on the data and their sources are reported in the Supplementary Information.

**2.3| Performance indices**

Performance indices were defined as described in **Tables 2-4** so that the higher the index, the higher the performance. Therefore, the best material should maximize its performance indices. The thermo-mechanical performance indices $M_1$ to $M_8$ were established following Ashby et al [23]. For the particulate level, index $E_7$, the quantity of particles released was normalized by human tolerance, which is of 1 for Ni, and 50 for $Al_2O_3$ [24]. The data were then converted to one hypothetic blade of dimensions $15 \times 5 \times 3 = 225$ cm$^3$. This would correspond to a CMSX-10 blade of $\sim 2$ kg and only 900 g for a blade made from NLA.

The price index $P_1$ and the rarity index $P_2$ were calculated using data from the literature, assuming a stable continuous production rate. Labor cost index $P_4$ and electricity cost index $P_5$ were evaluated based on a standard wage as in the USA with 16'000 USD per year at 40 hours per week and a rate of 0.15 USD per kWh, respectively.

The social impact of nickel and bauxite mining were calculated using data from reports that recorded the frequency of negative $f_-$ and positive $f_+$ social impact according to selected thematic areas [25–28]. The S parameters are the sum of these frequencies for each thematic area, multiplied by the amount of ore in reserve per country:

$$S_{x,NLA} = \sum_{country\ i} f_{theme\ j} \times m_{Bauxite\ reserve\ in\ country\ i}, \qquad \textbf{(eq 1)}$$

with $m$ in Mtons and $x$ the index number, and:

$$S_{x,superalloy} = \sum_{country\ i} f_{theme\ j} \times m_{Ni\ reserve\ in\ country\ i} \tag{eq 2}$$

**Table 2: Material performance indices.**

| Resistance against: | Index | Description |
|---|---|---|
| Creep | $M_1 = \dfrac{T}{n}$ | $T$: temperature (°C), $n$: Norton creep exponent |
| Thermal expansion | $M_2 = \dfrac{T}{CTE}$ | $T$: temperature (maximum, °C), $CTE$: coefficient of thermal expansion (K$^{-1}$) |
| Hot corrosion | $M_3 = \dfrac{T}{corr}$ | $T$: temperature (maximum, °C), $corr$: corrosion rate (µm.hr$^{-1}$) |
| Tension | $M_4 = \dfrac{\sigma_y}{\rho}$ | $\sigma_y$: yield stress (MPa), $\rho$: density (kg.m$^{-3}$) |
| Bending | $M_5 = \dfrac{\sigma_y^{2/3}}{\rho}$ | $\sigma_y$: yield stress (MPa), $\rho$: density (kg.m$^{-3}$) |
| Vibrations | $M_6 = \dfrac{\sqrt{E}}{\rho}$ | $E$: Young's modulus (MPa), $\rho$: density (kg.m$^{-3}$) |
| Catastrophic failure | $M_7 = \dfrac{K_{Ic}^2}{\sigma_y}$ | $K_{Ic}$: stress intensity factor (maximum, MPa.m$^{0.5}$), $\sigma_y$: yield stress (MPa) |
| High loads before fracture | $M_8 = \dfrac{K_{1c}}{\sigma_y}$ | $K_{1c}$: stress intensity factor (initial, MPa.m$^{0.5}$), $\sigma_y$: yield stress (MPa) |
| Fouling | $M_9 = \dfrac{\Pi_{superalloys}}{\Pi_{surface}}$ | $\Pi$: adhesion (cf SI equation S1) |
| Erosion | $M_{10} = \dfrac{1}{k_a} \cdot H$ | $H$: hardness (GPa), $k_a$: wear rate constant |

**Table 3: Environmental performance indices.** CTUh represents the Comparative Toxicity Unit for humans and the PM10 level corresponds to the quantity of particles that are of 10 µg/ m$^3$ or less.

| Not having a negative impact on: | Index | Description |
|---|---|---|

| Global warming | $E_1$ | (kg CO$_2$-eq/ blade)$^{-1}$ |
|---|---|---|
| Terrestrial acidification | $E_2$ | (kg SO$_2$-eq/ blade)$^{-1}$ |
| Energy consumption | $E_3$ | (MJ-eq/ blade)$^{-1}$ |
| Freshwater eutrophication | $E_4$ | (kg P-eq/ blade)$^{-1}$ |
| Human toxicity | $E_5$ | (CTUh/ blade)$^{-1}$ |
| Water usage | $E_6$ | (m$^3$ H$_2$O-eq/ blade)$^{-1}$ |
| Particulate levels | $E_7$ | (PM10 level µg/ blade)$^{-1}$ |

**Table 4: Socio-economic performance indices.**

| Not having a negative impact on: | Index | Description |
|---|---|---|
| Price | $P_1$ | (USD/blade)$^{-1}$ |
| Rarity | $P_2$ | (World reserves in kg/years left before exhaustion of resources)$^{-1}$ |
| Complexity | $P_3$ | (Number of steps)$^{-1}$ |
| Labor cost | $P_4$ | (Hours per blade * USD per hour)$^{-1}$ |
| Electricity cost | $P_5$ | (kWh/blade * USD/kWh)$^{-1}$ |
| Demography | $S_1$ | Gender imbalance (-), inflation (+), population growth (+) |
| Economy, income, security | $S_2$ | Bribery (-), Business (+), Income (+), Inequality (-), Poverty (-), Social tensions (-), Thefts and accidents (-) |
| Employment and education | $S_3$ | Child/ forced labor (-), Employment (+), Lack of freedom (-), Poor working conditions (-), Skills & education (+), Temporary jobs (-), Unemployment (-) |
| Environment, health and safety | $S_4$ | Environment impacts affecting health (-), Health impacts (-), Water use competition (-) |
| Human rights | $S_5$ | Cultural resources (-), Discrimination (-), Human rights (- |
| | | ), Indigenous rights (-), Lack of stakeholder inclusion (-) |
| Land use and territorial aspects | $S_6$ | Access to land (-), Expropriation/ displacement (-), Infrastructures (+) |
| Market | $S_7$ | % part of the global market |

## 3| Results

### 3.1| Materials performance

To investigate the potential of nacre-like alumina (NLA) as a replacement to Ni superalloys in the hot section of engines, we first compared their material performance using the indices previously defined (**Figure 4**). We also included conventional technical alumina ceramics and SiC/SiC CMCs to highlight the uniqueness of NLAs. Indices $M_1$ to $M_3$ characterize the resistance to the hot and corrosive atmospheres, indices $M_4$ to $M_8$ characterize the mechanical properties under the solicitations commonly found in turbine blades, and $M_9$ to $M_{10}$ describe the resistance to fouling and degradation.

**Figure 4A** confirms that ceramics systems present a higher resistance to high temperatures as compared to metallic materials, with little creep and corrosion. Alumina, however, is slightly less advantageous than SiC due to a larger thermal expansion. During turbine operation, a large thermal expansion might lead to the development of stresses during cyclic heating and cooling, causing friction or misfits. However, **Figure 4B,C** highlights the potential of NLAs to combine the thermal and chemical resistance of refractory ceramics with a mechanical performance tending towards Ni superalloys. This is well depicted in the radars charts of figure 4B,C, whereby the blue areas of NLAs overlap more extensively with the orange areas of the Ni superalloys, as compared to the other ceramic systems considered. Indeed, at room and elevated temperatures, bulk alumina and SiC/SiC CMCs behave very distinctly from Ni-superalloys. They exhibit higher performance in terms of resistance to tension, bending and vibrations but are less efficient to resist fracture. On the contrary, superalloys provide a high resistance to catastrophic failure and can withstand high loads until failure occurs. However, one particularity of NLAs is their toughness and resistance to catastrophic failure, thereby operating a shift of performance from the traditional ceramics towards metallic properties. Although NLA's resistance to tension and bending are decreased as compared to bulk alumina, the resistance to failure is augmented. Such property also occurs at high temperature which makes it particularly interesting for aircraft's turbine blades.

Along with thermo-mechanical properties, the usage of ceramics in lieu of Ni superalloys presents advantages in term of recoverable and unrecoverable deterioration [29]. We investigated this point by assessing the materials' resistance to fouling and erosion (**Figure 4D-F**). Fouling causes a drop in power output and efficiency of the engine due to contamination by airborne salts, industrial pollution particles, mineral deposits, etc. During fouling, particles present in the air impact the blades and stick to their surface depending on their temperature, viscosity, surface tension and wettability [30]. Alumina surfaces appear less prone to adhesion and fouling from a representative set of particulates considered, namely coal, sawdust, basalt and bitumen (**Figure 4D**). The resistance to fouling $M_\varsigma$ in wet environment was also higher than in dry environment for $Al_2O_3$ and SiC as compared to the metals. Since natural wet environment is generally located above seas and oceans and comprises salt, the resistance to fouling is particularly important to avoid corrosion. Furthermore, alumina surfaces are more hydrophilic than superalloys, assuming a similar surface roughness, with a contact angle of water on alumina 2 to 3 times lower than on superalloys. This is advantageous as it would facilitate maintenance and the washing of the blades to remove contaminants (**Figure 4E**). Finally, SiC and $Al_2O_3$ ceramics are more wear-resistant than Ni superalloys at temperatures up to 1000 °C (**Figure 4F**). The wear resistance of ceramics varies with grain size and orientation [31]. In the case of NLA, anisometric grains are oriented parallel to each other so that the surface parallel to the horizontal grains presents grain dimensions of around 10 µm. Perpendicularly to the grains, the surface displays grains of 200-300 nm width (see SI Figure S1). Therefore, in NLAs, the wear resistance parallel to the grain orientation is likely to be lower than for grains orientated perpendicularly. There are no data available to confirm the influence of elongated grain orientation on wear in NLAs but polymer-$Al_2O_3$ nacre-like composites were found to exhibit wear at least two times superior when the load is applied parallel to the grain as compared to perpendicular to the grain [32]. Furthermore, in nacre-like composites and ceramics, the hardness and mechanical properties are higher on vertically aligned grains, increasing the resistance to degradation along that direction [33]. In the hypothetical case of turbine blades made of NLA prepared using a pressing method, the surface would feature the basal plane of the grains corresponding to a horizontal alignment and likely similar wear resistance and hardness than standard alumina. However, these two properties could be increased if it is possible to orient the grains at the surface of the blades perpendicularly to this surface. For example, magnetically assisted slip casting is a method that demonstrated grain orientation control in NLAs compositions and could be used for this purpose [33,34]. Nevertheless, whether horizontal or vertical the grain orientation, the use of NLAs would still result in a higher erosion resistance as compared to the Ni-based superalloys.

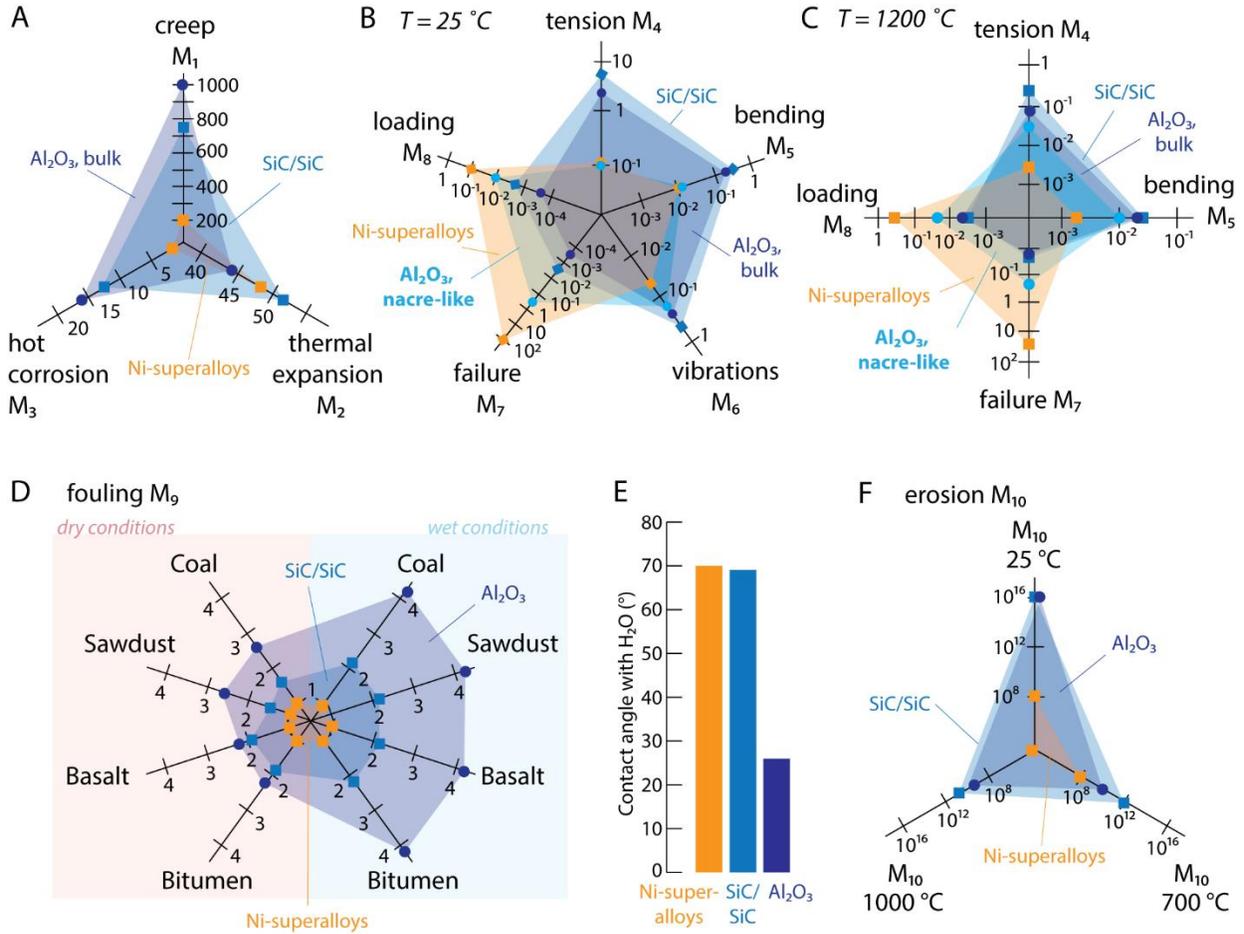

**Figure 4: Performance maps comparing Ni-based superalloys, SiC/SiC CMCs, alumina bulk and alumina nacre-like ceramics.** **(A)** Thermal properties, **(B,C)** mechanical properties at 25 and 1200 °C, **(D)** resistance to fouling in dry and wet conditions, **(E)** contact angle with water and **(F)** resistance to erosion at 25, 700 and 1200 °C.

Using a material with high resistance to corrosion and harsh environments could increase the lifetime of the engine and reduce the demand of maintenance. The mechanical performance of NLAs is promising. Simulation would be needed to estimate the stress level that these materials would need to withstand during a real engine operation.

## 3.2| Engine efficiency and payload

The efficiency of an engine is the product of its thermal efficiency $\eta_{th}$ by the propulsion efficiency $\eta_p$. In the following, we compare the efficiencies of an engine composed of Nickel superalloy only, or of NLAs only. Although the current engines employ ceramic-coated superalloys and NLAs engines do not exist, we base our comparison on the materials properties reported in the previous section. Following the Brayton cycle (**Figure 5A**), we calculated the efficiencies using the temperature at the compressor exit $T_b$ for a maximum work and a temperature at the combustor exit $T_c$ corresponding to the maximum working temperature of the material considered, using the following equations:

$$T_b = \sqrt{T_a T_c} \qquad (eq\ 3)$$

$$\eta_{th} = 1 - \frac{T_a}{T_b} \qquad (eq\ 4)$$

The resulting efficiencies at sea level ($T_a \sim 20°C$) and at cruise altitude of flight of 10 km ($T_a \sim -50°C$) were found to be significantly higher for an engine made of NLA, with an increase of 42% for the cruise altitude (**Figure 5B**). Furthermore, the propulsion efficiency is the ratio between the thrust power $\dot{W}_F$ and the power required to spin the propeller $\dot{W}_m$. Assimilating the propeller to a disc of radius $r$, we estimate the propulsion efficiency using:

$$\eta_p = \frac{\dot{W}_F}{\dot{W}_m} = \frac{\dot{W}_F}{\frac{1}{2}mr^2\varpi}, \qquad (eq\ 5)$$

where $m$ is the mass of the propeller, proportional to the mass of the blades, and $\varpi$ is the rotational speed. For a similar propeller volume and design, using NLA instead of superalloys would thus reduce the mass $m$ and increase $\eta_p$. Considering an Airbus A380, with 4 engines of type Trent 900, we estimate that the mass of an engine made entirely of CMSX-10 is of 6.5 tons, and only of 3 tons for a hypothetical engine made entirely of NLAs, *i.e.* ~54% lower. Such a decrease in weight would lead to an increase of about 116 % in the propulsion efficiency (**Figure 5C**).

Finally, a considerable advantage of using lightweight refractory materials in aircrafts is the overall decrease in aircraft weight. The Breguet's formula indicates the influence of the aerodynamics and structural parameters on the range $s$, the product of the time flight $t$ with its speed $V$ [35]:

$$s = V \cdot t = \frac{V\frac{L}{D}}{g\ SFC} \times \ln\left(1 + \frac{W_{fuel}}{W_{payload}+W_{structure}+W_{reserve}}\right), \quad \text{(eq 6)}$$

where $L$ is the lift, $D$ the drag, $g$ the gravitational constant, $SFC$ the specific fuel consumption, $W_{fuel}$ the weight of the fuel at the start, $W_{payload}$ the weight of the payload, $W_{structure}$ the weight of the structure and $W_{reserve}$ the weight of the reserve of fuel. Considering once more, a hypothetical Airbus A380 with 4 engines made of NLA, the total mass of the empty aircraft would be 5% lower than a A380 made Ni superalloys. Although a decrease in weight could lead to a larger increase in the range due to aerodynamics effects, a 5% decrease would also allow an increase of at least 5% in payload for the same amount of fuel and trip. Alternatively, it could also allow a longer flight, for the same payload and fuel, or a lower amount of fuel for the same flight distance and payload.

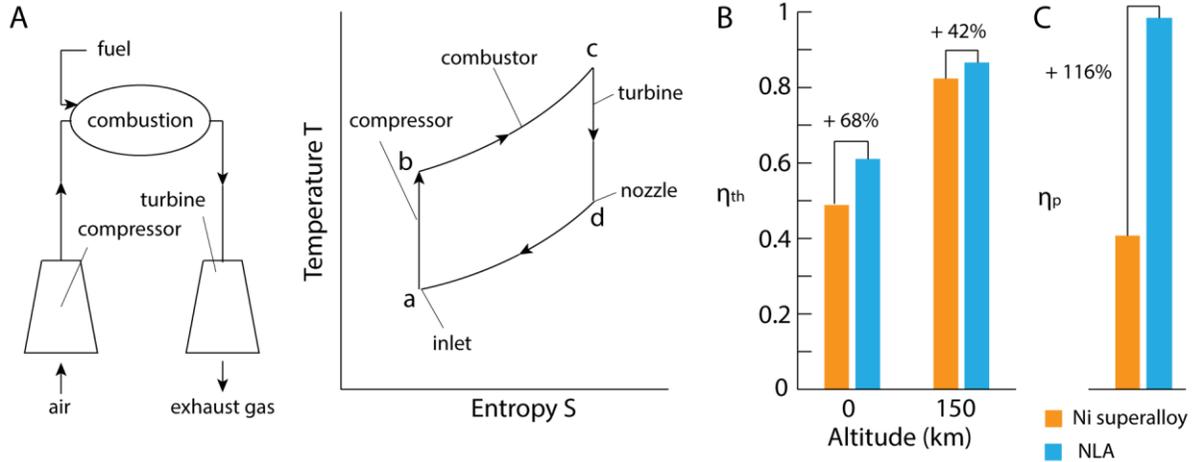

**Figure 5: Engine performance.** **(A)** Schematics of an engine and temperature-entropy graph of the Brayton cycle. **(B)** Thermal efficiency for two altitudes for an engine based on Ni superalloys and a hypothetical engine based on NLA. **(C)** Increase in the propulsion efficiency between the two types of engines.

A higher efficiency of the engines would also lead to a lower specific fuel consumption, SFC, which is the fuel consumption per unit thrust. According to Parker et al. [36], the improvements of the thermal and propulsion efficiencies would lead to at least 30% decrease in SFC per engine. For the same thrust, an aircraft consuming 802.3 kg/LTO such as the A320 could see its fuel consumption decrease to 561.6 kg/LTO with NLA's engines, with LTO the landing-take-off cycle [37]. Furthermore, with the price of kerosene of 0.84 USD/kg on average, the fuel price for one LTO would decrease from 671 USD to 469 [38]. This is a 30% decrease of fuel cost per flight. Finally, with $CO_2$ and $SO_2$ emissions being directly proportional to the fuel consumption [3], for one LTO, the $CO_2$ emissions could drop similarly by 30%, from 2527 to 1768, and the $SO_2$ emissions from 0.8 to 0.56.

The usage of NLA ceramics as replacement of nickel superalloys in aircraft engines has thus the potential to increase their efficiency, which may lead to increased profit due to the use of lesser fuel or a higher payload, as well as a reduction in the emission of greenhouse gases [35]. To better discuss the environmental and socio-economic impact of the usage of NLAs, the next section looks into the mining, fabrication, and end-of-life of NLA ceramics over Ni superalloys.

**3.3| Environmental, economic and social impacts of the production cycle**

**3.2.1| Mining**

The data on resource mining for the fabrication of the equivalent of one turbine blade indicate that alumina performs better on all collected factors, except terrestrial acidification and energy consumption (**Figure 6A**). Indeed, the extraction of metals and of alumina involves very different mechanisms. Alumina production, usually done *via* bauxite mining and the Bayer process, generates less $CO_2$, P and toxic compounds as compared to the mining of nickel, the major component of nickel superalloys. Also, the Bayer process uses less water and, although it releases quite a large amount of particulates, the tolerance level for these particulates is higher than for nickel. The main downside of the alumina process is the residual red mud produced [39]. The high alkalinity of this residue, with a pH ~ 13, strongly impacts living organisms if released in nature, as seen in 2010 in the Ajka disaster in Hungary [40]. This red mud is thus a threat to the environment. In comparison, ore refining and in particular, nickel ore refining, results in the loss of biodiversity by another mechanism that is the production of $SO_2$ which induces acidic rain. With more than 2 million tons of $SO_2$ released annually, large disasters have been observed, such as the destruction of 180,000 hectares of forests within 120 km of the Norilsk Nickel facility in Russia in 1992 [41].

In addition to the environmental impact during mining, the availability of the resources in the environment strongly determines their cost. For most of the metallic elements present in the CMSX-10 formulations, their rarity and price are greater than for alumina-based ceramics (**Figure 6B**). Several elements like Re and Ta are particularly rare and expensive, which is a problem since more ore is extracted every year, ultimately leading to their exhaustion. However, 3-6% of Re is necessary for reducing creep in Ni superalloys [42]. Ni is also getting depleted, with a prediction to run out of Ni by 2190 [43].

The evaluation of the social impact of mining has not been studied specifically for different ores. Instead, we draw a comparison based on the countries where mining of bauxite or of Ni are predominant. As a first approximation, we quantified the social impacts assuming countries having ore reserves will exploit them. Ore mining results in positive and negative impacts on the economy, income and security, employment and education, land use and territorial aspects, demography, environment, health and safety, and human rights of a country [25]. When new mining sites are deployed, there are reports of increase in demography, improved economy and employment but also, in rare cases, conflicts around the sites [25]. However, mining has been found to threaten the health, safety, education and income of the population when it is carried out without abiding regulations and respecting human rights [26,44]. We found that in general, there is not much difference in social impact between mining for nickel and mining for bauxite (**Figure 6C**). Yet, nickel mining poses a higher threat to human rights than bauxite mining whereas bauxite mining poses challenges on the local demography and land use. One important consideration is that the social impacts differ from region to region, with Africa and South America having a lesser positive mining impact than the other regions (**Figure 6D,E**). These social consequences are indeed correlated with the market demand, ore rarity and value, and governance.

Overall, despite remaining challenges, mining for bauxite has lower environmental consequences than mining for Ni, can be done at larger scale and lower cost, while having a similar social impact. In the event that bauxite production is to be further amplified, strict and clear governance must be established.

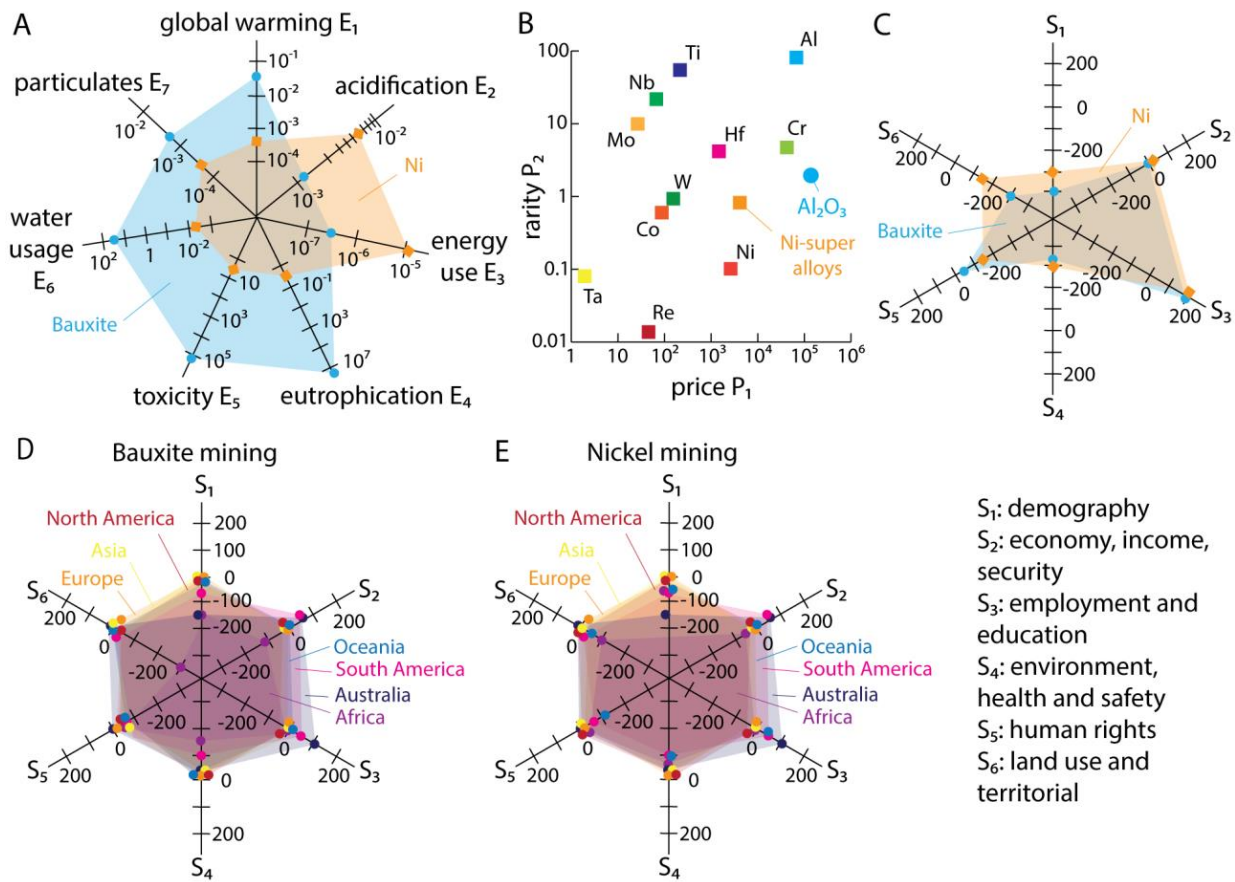

**Figure 6: Environmental and socio-economic impacts of mining.** (**A**) Environmental performance indices for mining of nickel and of bauxite. (**B**) Rarity factor as a function of the cost factor for the raw resources for nickel superalloy and alumina fabrication. Social impact

factors for the world for bauxite and nickel mining **(C)**, for continents for bauxite **(D)** and nickel **(E)** mining.

### 3.2.2| Fabrication

To compare the environmental impact for the production of the materials from raw resources, we considered conventional preparation of metallic materials by investment casting and the hot-pressing approach for NLA (**Figure 7**). Since NLAs prepared by hot-pressing cannot yet reach the shape complexity of a turbine blade, the comparison is done on the processing of a simple disc shape. Looking at the process from the ore to the final product, one main advantage of investment casting over hot-pressing is the lower number of steps: 3 for the superalloys as compared to 5 for the NLAs. However, investment casting uses a ceramic shell as a mold, therefore adding several steps to the process. Also, the fabrication of NLA requires the use of alumina particles with a platelet shape, whose synthesis may require a certain number of steps and chemicals (see SI Table S1). The environmental impacts of the two processes were thus evaluated following the simplified schemes depicted in **figure 7A**. Overall, the NLA processing consumes less energy and emits less $CO_2$ per blade as compared to the investment casting of superalloys (**Figure 7B**).

USA standard salary and electricity costs were used to assess the costs of the fabrication. Processing a disc made of NLA uses less electricity as compared to investment casting but consumes more time, thus leading to higher manpower cost (**Figure 7C**). In practice, manpower costs can be optimized by appropriate work and task management.

The social impact of fabrication has not been reported in the literature but discussion can be built upon the comparison of the % of the market share of the resources per countries vs % part of the aviation market (**Figure 7D**). It is noticeable that the countries poor in resources are sharing the highest part of the aviation market. Since the countries participating in aviation development are essentially located in Europe and North America, it suggests that these continents see positive social impacts. In future, the global repartition of the market share should be flattened between the countries to promote positive social and economic impacts in the countries possessing the raw resources as well.

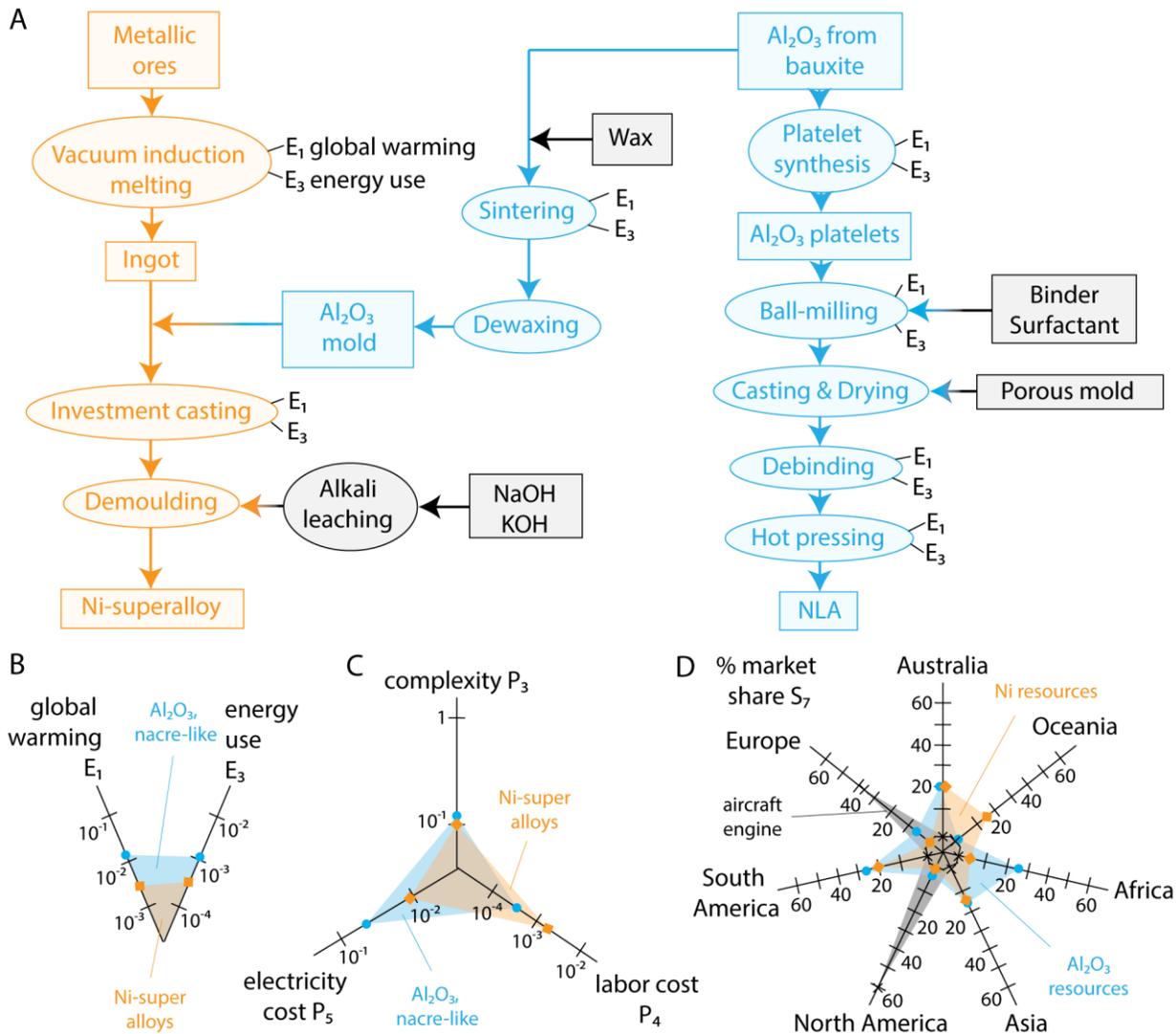

**Figure 7: Comparison of the fabrication processes of Ni-superalloys and NLAs. (A)** Workflow of the fabrication processes. **(B)** Environmental, **(C)** social and **(D)** economic impacts.

**3.2.3| End-of-life**

The end-of-life for current Ni-superalloys used in turbine blades is difficult to assess due to the lack of regulations [45]. Typically, materials that are not reused are either recycled, disposed of for landfilling, or burned to recover energy. In the case of aviation materials, the quantity of materials reused is difficult to estimate. At the end of their lifetime, aircrafts are stored on landsites in deserts for a few years before being dismantled. Parts like engines and turbines are generally reused [45].

Nickel superalloys can be recycled using pyrometallurgy, hydrometallurgy and pyro-hydrometallurgy methods [46]. Typically, after one recycling cycle, a CMSX-10 material can be recycled at 43.9% of the initial fabricated material [47]. However, metallic elements can be extracted from scrap at different rates depending on the method. For example, molten magnesium can be used to extract 97.2% of Ni from Ni-superalloys scrap [48], whereas only 62.32% can be extracted using a hydrometallurgical process involving leaching and sonication [49]. The hydrometallurgical process can recover up to 50% of Re, one of the rarest metal on Earth, as well as 58,5% of Cr, 60% of Co, 3.58% of Ta [49]. Another process based on electro-generated $Cl_2$ for leaching can extract up to 93% of Re [50]. Those recycling methods require the use of chemicals, electrical energy, water, whose specific environmental impact is yet to be determined. Finally, superalloys can also be grounded into powders and used as starting materials for creating new metallic parts, for example, via additive manufacturing [51]. On the contrary, alumina ceramics and refractories are rarely recycled and are most commonly disposed of in landfilling sites [52]. Recycling them generally consists in two strategies: grinding and reuse as cement additives or leaching for Al recovery. The addition of ground ceramic waste in Portland cements has shown increase in abrasion resistance and compression strength [53].

To approach the recycling potential of Ni-superalloys and ceramics, 2018 data from waste in the USA were used to quantify the percentage of waste from glass and ceramics versus metals to be recycled, in comparison to disposing in landfilling sites or burned for energy recovery (**Figure 8A**). Overall, glass and ceramics are more largely disposed in landfilling sites as compared to metals that are more often recycled. Since refractories like alumina are tougher to burn or leach due to their chemical and temperature resistance, it can be assumed that the % of landfilling will be even higher. The greenhouse gas emissions from these processes has also been reported [54,55] and show that refractories perform better in recycling and landfilling, whereas metals are more suited for reprocessing (**Figure 8B**). The poor performance of ceramics for reprocessing is due to the need for grinding and their use as additives to cements, which is a material with a large carbon footprint. Finally, the end-of-life of turbine parts may have considerable social impact in low-income countries where informal recycling and waste management by dumping are predominant [56].

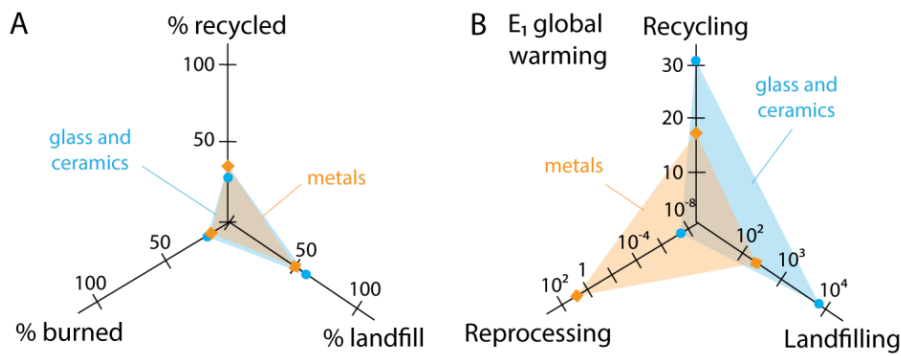

**Figure 8: End-of-life of metal and ceramic waste. (A)** % of Mton of waste recycled, burned or disposed in landfilling sites. **(B)** Global warming impact of those processes.

# 3| Discussion

## 3.1| Overall performance of NLA

Based on the performance indices gathered, NLAs present advantages on the engine performance and sustainability as compared to the nickel superalloys. Indeed, NLAs perform better mechanically than the superalloys on all properties at room and high temperatures, except for the failure-related properties. They also are less likely to be eroded or fouled which could significantly increase the engine performance. These properties would allow an easier cleaning of the blades and a longer lifetime. In parallel, the increased engine efficiency would lead to reductions in fuel consumption, generating profits and decreasing environmental impact. However, the main challenge remains on the safety and reliability of the material. Despite the significant improvement in resistance against catastrophic failure as compared to other ceramics, it is still much lower compared to superalloys. Furthermore, there is a large amount of thermomechanical properties that still need to be tested, such as resistance to thermomechanical fatigue and resistance to thermal and mechanical shocks. Further improvement on the toughness is also required for the development of reliable non-destructive testing and investigation methods to assure maintenance and routine check of the blades.

On the environment and socio-economic aspect, the use of alumina would make the blades cheaper due to the abundance of resources. This would lead to increased profits for the manufacturers. However, mining of bauxite should be carefully regulated and controlled in order to maintain a positive social impact, whereas the red mud byproduct still requires innovation to be managed. Furthermore, the fabrication of NLAs appears greener than investment casting. Investment casting makes use of both metallurgy and ceramic processes whereas the process of NLAs only relies on ceramic paths. One major issue to be overcome is the process of fabricating

NLAs with complex shapes and controllable angles of the brick-and-mortar architecture. As of current technologies, hot-pressing methods have been used to densify the brick-and-mortar structures, thereby leading to laminates. Near net-shape processing methods relying on layer-by-layer assembly, abnormal grain growth, and *in situ* mineralization have yet to yield similar properties as the hot pressed NLAs [34,57,58]. The hypothesis put forward is that the interface properties between the grains should remain weak enough for intergranular crack propagation [59]. Finally, the low recyclability of refractory might pose challenges with regards to their disposal. One last limitation of current NLAs is the lack of repair capabilities. In contrast, research on Ni-superalloys repair has led to promising strategies (see SI Table S2).

Overall, the use of NLAs for turbine blade applications is promising if their reliability is increased and if there is complexified manufacturing methods to allow for turbine blade shapes. Furthermore, global regulation and strategies to ensure proper resource mining and waste management should be developed and implemented. The following section shows the most recent advances towards increased reliability and complex manufacturing.

**3.2| Latest advances on toughening and shaping NLAs**

To tackle the issue of reliability in NLAs, we explore the potential use of ceramic-polymer, metal and ceramic composites (**Figure 9**). The map of the thermomechanical performance based on performance indices at room temperature shows that polymer-based NLAs are not suitable for high temperature applications but alumina-metal composites are able to combine good temperature properties with excellent toughness. This is particularly true of the $Al_2O_3$-Zr glass composites prepared by spark plasma sintering (SPS) [60]. Another very interesting solution utilizes a co-extrusion method with a composition based on $Al_2O_3$ and Ni [61,62]. The co-extrusion of slurries through a nozzle resembles additive manufacturing and therefore has the potential to achieve not only outstanding performance but also shape complexity. However, besides a need to complete the mechanical characterization with testing at higher temperatures and under cyclic loads, the usage of a ceramic-metal composite system would also increase the mining and recycling environmental impact and cost. Thus, there is a trade-off to find between performance and complexity of fabrication and recycling.

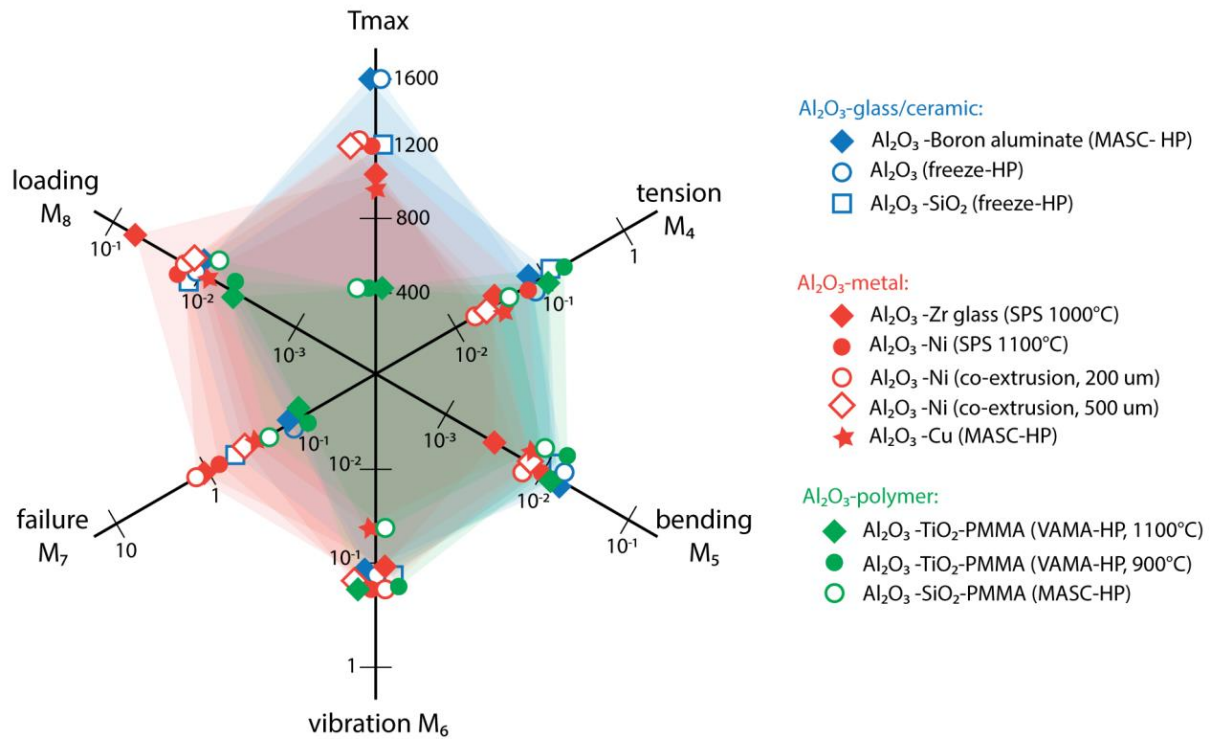

**Figure 9: Mechanical performance indices for nacre-like composites.** The mortars are glass or ceramic (blue) [22,63,64], metallic (red) [62,63,65,66] and organic (green) [63,67]. MASC refers to magnetically assisted slip casting, HP to hot pressing and freeze to freeze-casting. SPS is spark plasma sintering, VAMA vacuum-assisted slip casting and PMMA polymethyl methacrylate.

**4| Conclusions**

We have compared the thermomechanical performance, environmental and socio-economic of nickel superalloys with NLAs for application as turbine blades in aircraft engines. NLAs are promising in terms of engine efficiency, reduced fuel consumption, abundance of resources available and cheaper cost. However, mechanical reliability, shaping and recycling are still major problems hindering their implementations in the industry and in particular, as turbine blades for aviation. Several recent developments have explored ceramic-metal composites and the usage of additive manufacturing to solve those challenges. However, thermomechanical characterization and life cycle assessment are yet to be completed. Further studies will also need to consider the assembly of those materials, testing via non-destructive means, and maintenance. Lastly, the use of cheaper materials for more efficient flight may lead to the problem of increased air traffic worldwide, ultimately leading to increased emissions and pollution. Comprehensive overview of the field and of the goals set to build a sustainable aviation domain will thus need to be considered if NLAs make their way into industrial production.

**Supporting information.**

Supporting information containing details and the raw data collections are available.

**Acknowledgment**

The authors acknowledge funding from the National Research Foundation, Singapore, with the Fellowship NRFF12-2020-0006.

**Author contributions**

JSC and HLF gathered and analyzed the data. HLF wrote the manuscript. All authors discussed the results and reviewed the manuscript.

**Competing interests**

The authors declare no competing interests.